\documentclass[pra,twocolumn,superscriptaddress,10pt]{revtex4-1}
\usepackage{times,epsfig,amssymb,amsfonts,amsmath,graphicx}
\usepackage{graphicx,epsfig,epstopdf}
\usepackage[dvipsnames]{xcolor}
\usepackage{amsmath}
\usepackage{subfigure}
\usepackage{mathrsfs}
\usepackage{bm}
\usepackage{times}
\usepackage{color}%

\begin{document}

\title{Monogamy of quantum entanglement}
\author{Xiao-Lan Zong}
\affiliation{Universities Joint Key Laboratory of Photoelectric Detection Science and Technology in Anhui
Province, and School of Physics and Materials Engineering, Hefei Normal University,
Hefei 230601, China}

\author{Hao-Hao Yin}
\affiliation{Universities Joint Key Laboratory of Photoelectric Detection Science and Technology in Anhui
Province, and School of Physics and Materials Engineering, Hefei Normal University,
Hefei 230601, China}

\author{Wei Song}\email{wsong315@qq.com}
\affiliation{Universities Joint Key Laboratory of Photoelectric Detection Science and Technology in Anhui
Province, and School of Physics and Materials Engineering, Hefei Normal University,
Hefei 230601, China}

\author{Zhuo-Liang Cao}
\affiliation{Universities Joint Key Laboratory of Photoelectric Detection Science and Technology in Anhui
Province, and School of Physics and Materials Engineering, Hefei Normal University,
Hefei 230601, China}

\date{\today}

\begin{abstract}
Unlike classical correlation, quantum entanglement cannot be freely shared among many parties. This restricted shareability of entanglement among multi-party systems is known as monogamy of entanglement, which is one of the most fundamental properties of entanglement. Here, we summarize recent theoretical progress in the field of monogamy of entanglement. We firstly review the standard CKW-type monogamy inequalities in terms of various entanglement measures. In particular, the squashed entanglement and one-way distillable entanglement are monogamous for arbitrary dimensional systems. We then introduce some generalized version of monogamy inequalities which extend and sharpen the traditional ones. We also consider the dual polygamy inequalities for multi-party systems. Moreover, we present two new definitions to define monogamy of entanglement. Finally, some challenges and future directions for monogamy of entanglement are highlighted.
\end{abstract}

\maketitle

\section{introduction}

Quantum entanglement has been recognized as the most important resource in many quantum information processing tasks\cite{Bennett1993, Bennett1992, Gisin2002}. One of the essential differences between quantum entanglement and classical correlation is that quantum entanglement cannot be freely shared among many parties. For example, in a multi-party state, if two parties are maximally entangled, then none of them can share entanglement with any part of the rest of the system. This restriction of entanglement shareability among multi-party systems is known as the monogamy of entanglement (MOE)\cite{Terhal:2004, Yang:2006}.

Since the MOE restricts on the amount of information that an eavesdropper could potentially obtain the secret key extraction, it is a crucial property that guarantees quantum key distribution secure\cite{Gisin2002,Terhal:2004,Pawlowski:2010}. MOE also has many fundamental applications in other areas of physics, including classification of quantum states\cite{Dur:2000,Giorgi:2011,Prabhu:2012}, no-signaling theories\cite{Streltsov:2012}, condensed-matter physics\cite{Ma:2011,Brandao:2013, Artur:2013}, statistical physics \cite{Bennett:2014} and even black-hole physics \cite{Lloyd:2014}.

An important basic question in the study of MOE is to determine whether a given entanglement measure is monogamous. Usually, there are several ways to define the monogamy property of entanglement measure. Originally, a monogamy relation of entanglement measure $E$ is quantitatively displayed as an inequality of the following form
\begin{eqnarray}
E(\rho_{A|BC})\geq E(\rho_{A|B})+E(\rho_{A|C})
\label{Eq1}
\end{eqnarray}
where $E(\rho_{A|BC})$ is an entanglement measure quantifying the degree of entanglement between subsystems A and BC, and $E(\rho_{A|B})$ ($E(\rho_{A|C})$) is the bipartite entanglement between A and B (A and C)(See Fig.1 for a graphical representation). This inequality means that the sum of entanglement between A and each of the other parties B or C cannot exceed the entanglement between A and BC. Using squared concurrence(SC) as an entanglement measure, Coffman, Kundu and Wootters (CKW) proved the first monogamy inequality for three qubit states\cite{Coffman:2000} which we shall refer to as the CKW inequality. The CKW inequality was later generalized by Osborne and Verstraete for arbitrary multi-qubit system. It should be noticed that the entanglement of formation(EOF), when not squared, does not obey the monogamy relation given by Eq.(\ref{Eq1})\cite{Fanchini:2013}. Besides SC, it was further proven that similar monogamy inequality can be established for the squared entanglement of formation(SEF)\cite{Baiprl:2014,Baipra:2014}, R\'{e}nyi-$\alpha$ entanglement(R$\alpha$E)\cite{Kim:2010}, the squared R\'{e}nyi-$\alpha$ entanglement(SR$\alpha$E)\cite{Song:2016}, Tsallis-$q$ entanglement(TqE)\cite{Kimpra:2010}, the squared Tsallis-q entanglement(STqE)\cite{Luo:2016,Yuan:2016}, and unified-($q,s$) entanglement\cite{Kim:2011}. The establishment of these inequalities depends on monogamy inequality of SC. In this sense, these inequalities can be classified into concurrence-based monogamy relations. For high-dimensional systems, it has been shown that monogamy inequality of SC can be violated due to the existence of counterexamples\cite{Ou:2007,Kim:2009}. At present, it is still unclear whether other concurrence-based monogamy relations hold in high-dimensional systems.

Another way to generalize the CKW inequality is using negativity\cite{Ou:2007b} or convex-roof extended negativity (CREN)\cite{Kim:2009}, and CREN is a good candidate for MOE without any known example violating its CKW-type inequality even in higher-dimensional systems\cite{Kim:2009}. More recently, Gao \emph{et al} \cite{Gao:2020} established a class of CKW-type monogamy inequalities based on the $\mu$-th power of logarithmic negativity and logarithmic convex-roof extended negativity(LCREN). The CKW-type inequality was also generalized to other entanglement measures, such as squashed entanglement\cite{Koashi:2004,Yang:2009}, one-way distillable entanglement\cite{Koashi:2004} and continuous-variable entanglement\cite{Adesso:2006,Hiroshima:2007,Adesso:2007}. Among them, the squashed entanglement and one-way distillable entanglement fulfill Eq.(\ref{Eq1}) for arbitrary dimensional systems. Furthermore, other types of monogamy relations were presented in Ref.\cite{Bandyopadhyay:2007,Yu:2009,Cornelio:2013,Oliveira:2014,Zhu:2014,Zhu:2015,Liu:2015,Jin:2017,Jin:2018,Kimc:2018,Gao:2018,Char:2020,Shi:2020,Yang:2021,Shi:2021,Yang:2018,Eltschka:2009}. In particular, Regula \emph{et al}\cite{Regula:2014,Regula:2016} have proposed a set of strong monogamy(SM) inequalities sharpening the conventional CKW-type inequality. For the validity of SM inequality, an extensive numerical evidence was presented for four qubit pure states together with analytical proof for some cases of multi-qubit systems.

On the other hand, the polygamous property can be regarded as another kind of entanglement constraints in multi-qubit systems, and Gour \emph{et al}\cite{Gour:2005} established the first dual polygamy inequality for multi-qubit systems using concurrence of Assistance(CoA). Subsequently, polygamy inequalities was generalized into various entanglement measures\cite{Buscemi:2009,Jin:2019,Kimb:2012,Kim:2019,Lai:2021,Song:2019,Kim:2016,Song:2017,Kim:2012,Kim:2018}.

However, the main problem with the definition of monogamy in Eq.(\ref{Eq1}) is that their validity is not universal, but depends on the specific choice of $E$. Moreover, several important measures of entanglement do not satisfy the relation (\ref{Eq1}). Therefore, the summation in the right-hand sides of Eq.(\ref{Eq1}) is only a convenient choice and not a necessity. To overcome this problem, one attempt is to replace Eq.(\ref{Eq1}) with a family of monogamy relations of the form $E(\rho_{A|BC})\geq f(E(\rho_{A|B}),E(\rho_{A|C}))$, where $f$ is some function of two variables that satisfies certain conditions\cite{Lancien:2016}. Another approach is based on the definition of monogamy relations without inequalities introduced in Ref.\cite{Gour:2018}. According to this definition, we can reproduce the traditional monogamy relations similar to Eq.(\ref{Eq1}) by replacing $E$ with $E^{\alpha}$ for some $\alpha>0$.

\begin{figure}[htb]
\includegraphics[scale=0.2,angle=0]{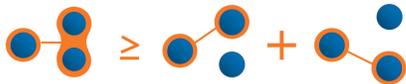}
\caption{(color online). Schematic picture of the CKW-type monogamy relation described by Eq.(1).}\label{fig1}
\end{figure}

In this review, we focus on introducing theoretical advances on monogamy of quantum entanglement but not include the topic of quantum correlations, see Ref.\cite{Dhar:2017} for the summary of recent advances in monogamy of quantum correlations. In Sec.II, we firstly review the standard CKW-type monogamy inequalities in terms of various entanglement measures. In Sec.III, we then introduce some other types of monogamy inequalities which extend and sharpen the existing ones. In Sec.IV, we focus on reviewing the dual polygamy inequalities for multi-qubit systems. The new definitions of MOE are discussed in Sec.V. Finally, in Sec.VI, we give some concluding remarks.

\section{CKW-type inequalities}

In this section we briefly review the CKW-type monogamy inequality and we divide them into three categories according to different entanglement measures.

\subsection{Concurrence-based inequalities}

We start by recalling the monogamy inequality introduced by Coffman, Kundu and Wootters(CKW)\cite{Coffman:2000} for three-qubit states

\begin{eqnarray}
C^2(\rho_{A|BC})\geq C^2(\rho_{A|B})+C^2(\rho_{A|C})
\label{Eq2}
\end{eqnarray}
where $C^2$ denote the squared concurrence for quantifying bipartite entanglement. For an arbitrary two-qubit state, concurrence is defined as\cite{Hill:1997,Wootters:1998} $C(\rho)=max\{0,\lambda_1-\lambda_2-\lambda_3-\lambda_4\}$, in which ${\lambda_1},{\lambda_2},{\lambda_3},{\lambda_4}$ are the square root of the eigenvalues of the matrix $\rho({\sigma_y}\otimes{\sigma_y})\rho^{*}({\sigma_y}\otimes{\sigma_y})$in decreasing order, ${\sigma_y}$ is the Pauli spin matrix and $\rho^{*}$ denotes the complex conjugate of $\rho$. Usually, Eq.(\ref{Eq2}) is termed as CKW inequality, and it shows a tradeoff relation between the amount of entanglement shared by qubits A and B and the entanglement shared by qubits A and C. For three-qubit pure states, the difference between left and right-hand sides of Eq.(\ref{Eq2}) is interpreted as a genuine three-qubit entanglement measure, three tangle. It has been proved that three-tangle is an entanglement monotone, and the generalization of the three-tangle to mixed states can be obtained by the convex roof method\cite{Uhlmann:2000,Lohmayer:2006,Eltschka:2008}. Later, CKW inequality was generalized to the multi-qubit case, $C^2(\rho_{A|{B_1}...{B_{n-1}}})\geq C^2(\rho_{A|{B_1}})+\cdots+C^2(\rho_{A|{B_{n-1}}})$, in which $C^2(\rho_{A|{B_1}...{B_{n-1}}})$ quantifies bipartite entanglement in the partition $A|{B_1}...{B_{n-1}}$, and $C^2(\rho_{A|{B_i}})$ characterizes the two-qubit entanglement with $i=1,2,...,{n-1}$. Unfortunately, the CKW inequality is violated if we use EOF instead of SC. In order to obtain a similar monogamy inequality, Bai \emph{et al}\cite{Baiprl:2014} proved that the squared entanglement of formation (SEF) obeys the CKW-type monogamy relation for an arbitrary multi-qubit mixed state. Based on this new monogamy relation, they further constructed entanglement indicators which detect genuine multiqubit entanglement even in the case of three-tangle being zero. Another generalization is using  R$\alpha$E which is a well-defined entanglement measure introduced in Ref.\cite{Kim:2010}. For a bipartite pure state $\left|\psi\right\rangle _{AB}$, the R$\alpha$E is defined as

\begin{eqnarray}
E_{\alpha}(|\psi_{AB}\rangle):= S_\alpha(\rho_A) = \frac{1}{1-\alpha}\log_2(\mbox{tr}\rho_A^\alpha)
\label{Eq3}
\end{eqnarray}
where the R\'{e}nyi-$\alpha$ entropy is $S_\alpha(\rho_A)=[\log _2(\sum_i\lambda_i^{\alpha})]/(1-\alpha)$ with
$\alpha$ being a nonnegative real number and $\lambda_i$ being the eigenvalue of reduced density matrix
$\rho_A$. The R\'{e}nyi-$\alpha$ entropy $S_\alpha \left( \rho  \right)$ converges to the von Neumann
entropy when the order $\alpha$ tends to 1. For a bipartite mixed state $\rho_{AB}$, the R$\alpha$E
is defined via the convex-roof extension $E_\alpha(\rho _{AB})=\mbox{min} \sum_i p_i E_\alpha(|{\psi _i }_{AB}\rangle)$,
where the minimum is taken over all possible pure state decompositions of ${\rho_{AB}=\sum\limits_i{p_i
\left| {\psi _i } \right\rangle _{AB} \left\langle {\psi _i } \right|} }$. It is shown that R$\alpha$E obeys the CKW-type inequality for $\alpha\geq2$, but
this monogamy relation does not cover the case of EOF, which corresponds to R$\alpha$E with the order $\alpha=1$. Subsequently, Song \emph{et al}\cite{Song:2016} proved that the SR$\alpha$E with the order $\alpha\geq{\sqrt{7}-1}/2\simeq0.823$ obeys a general monogamy relation in an arbitrary multi-qubit mixed state. This result provides a broad class of monogamy inequalities including the monogamy relation of the SEF as a special case. Recently, CKW-type inequalities in terms of TqE, STqE and unified-($q,s$) entanglement for arbitrary multi-qubit mixed state have also been proved in\cite{Kimpra:2010,Luo:2016,Yuan:2016,Kim:2011}. The above discussed monogamy inequalities are termed as concurrence-based inequalities since their validity are conditioned on the truth of the monogamy inequality of SC. Moreover, it has been shown that the $\mu$ th$(\mu\geq2)$ power of concurrence and the $\mu$ th$(\mu\geq\sqrt{2})$ power of EOF satisfy the monogamy inequalities, respectively\cite{Zhu:2014}. In addition, Kumar showed in Ref.\cite{Kumar:2016} that monogamy is preserved for raising the power and polygamy is maintained for lowering the power, and this result has also been pointed our in the earlier paper\cite{Salini:2014}. The CKW inequality is invalid for higher-dimensional systems due to the existence of counterexamples for states in the systems $3\otimes3\otimes3$\cite{Ou:2007} and $3\otimes2\otimes2$\cite{Kim:2009}. It is still an open problem yet to be answered whether other concurrence-based monogamy relations hold in high-dimensional systems since the exact formula for these cases are missing.

\subsection{Negativity-based inequalities}

Another well-known bipartite entanglement measure is negativity\cite{Vidal:2002,Plenio:2005}. It is a rare entanglement measure which is easy to compute for pure as well for mixed bipartite states. For any bipartite state $\rho_{AB}$ in the Hilbert space $\mathcal{H}_A\otimes\mathcal{H}_B$, the negativity is defined by

\begin{eqnarray}
\mathcal{N}(\rho_{AB})=\frac{\|\rho_{AB}^{T_A}\|-1}{2}
\label{Eq4}
\end{eqnarray}
where $\rho_{AB}^{T_A}$ is the partially transposed matrix of $\rho_{AB}$ with respect to the subsystem A, $ \|X\|=Tr\sqrt{XX^\dag}$ denotes the trace norm of $X$. In order for any maximally entangled state in $2\otimes2$ systems to have the negativity one, we use the following definition of negativity: $\mathcal{N}(\rho_{AB})=\|\rho_{AB}^{T_A}\|-1$. It has been shown that for any pure three-qubit state, the squared negativity satisfies the following CKW-type monogamy inequality\cite{Ou:2007b}

\begin{eqnarray}
\mathcal{N}^2(\rho_{A|BC})\geq \mathcal{N}^2(\rho_{A|B})+\mathcal{N}^2(\rho_{A|C})
\label{Eq5}
\end{eqnarray}
where $\mathcal{N}^2(\rho_{A|B})$ and $\mathcal{N}^2(\rho_{A|C})$ are the negativities of the mixed states
$\rho_{AB}$ and $\rho_{AC}$, respectively. For any $n$-qubit pure states, the $\mu$-th($\mu\geq2$)power of negativity satisfies the monogamy inequality\cite{He:2015}: $\mathcal{N}^{\mu}_{A|{B_1\ldots}B_{n-1}}(|\psi\rangle)\geq \mathcal{N}^{\mu}_{A|{B_1}}+\cdots+ \mathcal{N}^{\mu}_{A|{B_{n-1}}}$. The definition in Eq.(\ref{Eq3}) cannot distinguish positive partial transposition(PPT) bound entangled states\cite{Peres:1996,Horodecki:1996,Horodecki:1998} from separable states, and for a bipartite mixed state, its convex roof extended negativity(CREN) is modified as\cite{Kim:2009}

\begin{eqnarray}
\mathcal{N}_c(\rho_{A|B})=min\sum\limits_k{p_k\mathcal{N}(|\phi_k\rangle_{A|B})}
\label{Eq6}
\end{eqnarray}
where the minimum is taken over all possible pure state decompositions of $\rho_{AB}=\sum\limits_k{p_k|\phi_k\rangle_{AB}\langle\phi_k|}$. CREN provides a perfect discrimination of PPT bound entangled states and separable states in any bipartite quantum system. For an arbitrary $n$-qubit state $\rho_{AB_1\ldots B_{n-1}}$, the square of CREN satisfies the following monogamy inequality: $\mathcal{N}_c^2(\rho_{A|B_1\ldots B_{n-1}})\geq \mathcal{N}_c^2(\rho_{A|B_1})+\cdots+ \mathcal{N}_c^2(\rho_{A|B_{n-1}})$. This inequality still holds for the counterexamples that violate CKW inequality in higher dimensional systems. Further generalization for the $\mu$-th power of CREN has been shown in Ref.\cite{Luob:2015}. Recently, Gao \emph{et al}\cite{Gao:2020} generalized the concept of logarithmic negativity\cite{Plenio:2005} to logarithmic convex roof extended negativity(LCREN). For any bipartite state $\rho_{AB}$, LCREN is defined as
\begin{eqnarray}
E_{\mathcal{N}_c}(\rho_{AB})=log_2[{\mathcal{N}_c}(\rho_{AB})+1]
\label{Eq7}
\end{eqnarray}
and Gao \emph{et al} have shown that LCREN is an entanglement monotone under LOCC operations but not convex. For any $n$-qubit pure state $|\psi\rangle_{A{B_1\ldots}B_{n-1}}$, the $\mu$-th power of logarithmic negativity obeys the CKW-type inequality $E_\mathcal{N}^{\mu}(\rho_{A|B_1\ldots B_{n-1}})\geq E_\mathcal{N}^{\mu}(\rho_{A|B_1})+\cdots+ E_\mathcal{N}^{\mu}(\rho_{A|B_{n-1}})$ for $\mu\geq4\sqrt{2}$, and similar monogamy inequality also holds for arbitrary $n$-qubit state $\rho_{AB_1\ldots B_{n-1}}$ in terms of LCREN. These results indicate that entanglement measure without convexity can also obey the monogamy inequality.

\subsection{Other CKW-type inequalities}

We now summarize other CKW-type inequalities in terms of various entanglement measure. Firstly, we consider the squashed entanglement introduced in Ref.\cite{Tucci:2002,Christandl:2004}, which is the first additive measure with good asymptotic properties. It is defined as
\begin{eqnarray}
E_{sq}(\rho_{AB})= inf\{\frac{1}{2}I(A:B|E): \rho_{AB}=Tr_E(\rho_{ABE})\}
\label{Eq8}
\end{eqnarray}
where the infimum is taken over all extensions $\rho_{ABE}$ of the state $\rho_{AB}$ and $I(A:B|E)=S(\rho_{AE})+S(\rho_{BE})-S(\rho_{ABE})-S(\rho_{E})$ is the conditional quantum mutual information. For any tripartite state $\rho_{ABC}$, Koashi and Winter\cite{Koashi:2004} have proved that squashed entanglement obeys the following CKW-type inequality:

\begin{eqnarray}
E_{sq}(\rho_{A|BC})\geq E_{sq}(\rho_{A|B})+E_{sq}(\rho_{A|C})
\label{Eq9}
\end{eqnarray}
and the above form of inequality is also true for the one-way distillable entanglement introduced in Ref.\cite{Koashi:2004}. Although squashed entanglement and one-way distillable entanglement satisfies the CKW inequality for arbitrary dimensional systems, there is no analytical formula to calculate these entanglement measures.

The CKW-type inequality has also been generalized to the continuous variable systems. By introducing the continuous-variable (CV) tangle (contangle) to quantify entanglement
sharing in Gaussian states, Adesso \emph{et al} \cite{Adesso:2006}proved the monogamy inequality for arbitrary three-mode Gaussian states and for symmetric $n$-mode Gaussian states. Here, contangle is defined as the convex roof of the square of the logarithmic negativity. Moreover, Hiroshima \emph{et al} have generalized the monogamy inequality to all $n$-mode Gaussian states of in terms of squared negativity\cite{Hiroshima:2007}.

\section{Strong monogamy inequalities}
In this section we focus on reviewing some generalized version of monogamy relation. It is well known that tightening the monogamy inequalities can provide a precise characterization of the entanglement sharing and distribution in multipartite systems, thus it is important to find tight monogamy inequalities for various entanglement measure. We first consider the strong monogamy(SM) inequality introduced by Regula \emph{et al}\cite{Regula:2014}. For an $n$-qubit pure state $|\psi\rangle$, it was conjectured that the following inequality holds:

\begin{eqnarray}
\tau(|\psi\rangle_{A_1|A_2\ldots A_n})\geq \sum_{m=2}^{n-1}\sum_{\vec{j}^m}\tau({\rho_{A_1|A_{j_1^m}|\ldots|A_{j_{m-1}^m}}})^{\frac{m}{2}}
\label{Eq10}
\end{eqnarray}
where the index vector $\vec{j}^m=(j_1^m,\ldots,j_{m-1}^m)$ spans all the ordered subsets of the index set $\{2,\ldots,n\}$ with $m-1$ distinct elements, and $\tau({\rho_{A_1|A_{j_1^m}|\ldots|A_{j_{m-1}^m}}})$ is defined as
\begin{eqnarray}
&\tau&({\rho_{A_1|A_{j_1^m}|\ldots|A_{j_{m-1}^m}}})\nonumber\\
&=&[\min_{\{p_h,|\psi_h\rangle\}}\sum_h{p_h\sqrt{\tau({|\psi_h\rangle_{A_1|A_{j_1^m}|\ldots|A_{j_{m-1}^m}}})}}]^2
\label{Eq11}
\end{eqnarray}
with the minimization over all possible pure state decompositions $\rho_{A_1A_{j_1^m}\ldots A_{j_{m-1}^m}}=\sum_{h} p_h|\psi_h\rangle_{A_1A_{j_1^m}\ldots A_{j_{m-1}^m}}\langle\psi_h|$. The right side of Eq.(\ref{Eq10}) appears in between the both side of the $n$-qubit CKW inequality, therefore it is a stronger inequality. The difference between left and right hand side of Eq.(\ref{Eq10}) is defined as $n$-tangle which is a quantifier of genuinely entanglement shared among $n$-partites. In fact, this inequality comes from the strong monogamy inequality of continuous variable Gaussian states introduced in Ref.\cite{Adesso:2007}. Eq.(\ref{Eq10}) reduces to normal three-qubit CKW inequality for $n=3$. For a four-qubit state $|\psi\rangle$, the SM inequality can be written as: $\tau_{A_1|A_2A_3A_4}\geq \tau^{(2)}_{A_1|A_2}+\tau^{(2)}_{A_1|A_3}+\tau^{(2)}_{A_1|A_4}+[\tau^{(3)}_{A_1|A_2|A_3}]^{3/2}+[\tau^{(3)}_{A_1|A_3|A_4}]^{3/2}+[\tau^{(3)}_{A_1|A_2|A_4}]^{3/2}$. For the validity of SM inequality, an extensive numerical evidence has been presented for four-qubit state together with analytical proof for some cases of multi-qubit state. Another generalization of SM inequality in terms of squared convex roof extended negativity(SCREN) has been presented by Choi and Kim\cite{Choi:2015}, and it is shown that the superposition of the generalized W-class states and vacuum (GWV) states satisfy the SM inequality based on SCREN. In Ref.\cite{Kimb:2016}, Kim further proved that SM inequality holds good even in a class of higher dimensional state where the original SM inequality fails.

Next we present some other generalized version of monogamy relation. In Ref.\cite{Jin:2017}, Jin \emph{et al} have investigated tighter entanglement monogamy relations related to $C^{\mu}$ and $E^{\mu}$ for ${\mu}\geq2$ and ${\mu}\geq\sqrt{2}$, respectively. Using the Hamming weight
of the binary vector related with the distribution of subsystems, Kim\cite{Kimc:2018,Gao:2018} established a class of monogamy inequalities of multi-qubit entanglement based on the $\mu$-th power of unified-($q, s$) entanglement. Other approaches to construct tighter monogamy inequalities in terms of various entanglement measures were also proposed in Ref.\cite{Jin:2018}. Moreover, Oliveira \emph{et al}\cite{Oliveira:2014} proposed a monogamy relation in the linear version for a three-qubit system, which was proved by Liu \emph{et al}\cite{Liu:2015}. In. Ref.\cite{Shi:2021}, Shi \emph{et al} generalized the multi-linear monogamy relation for a multi-qubit system in terms of EOF and concurrence.

\section{Polygamy inequalities}
In previous section, we have reviewed MOE which reveals the limited shareability of multiparty quantum entanglement, the assisted entanglement was shown to have a dually monogamous property in multi-party quantum systems, i.e, polygamy of entanglement(PoE). PoE is mathematically characterized as the polygamy inequality

\begin{eqnarray}
E_a(\rho_{A|BC})\leq E_a(\rho_{A|B})+E_a(\rho_{A|C})
\label{Eq12}
\end{eqnarray}
for a three-party quantum state and $E_a(\rho_{A|BC})$ denotes the bipartite assisted entanglement in the partition $A|BC$. In contrast to monogamy inequality, which provides an upper bound on the bipartite shareability of entanglement in multi-party systems, the polygamy inequality in Eq.(\ref{Eq12}) provides a lower bound for distribution of bipartite entanglement in multi-party systems.

The polygamy inequality in (\ref{Eq12}) was first proposed in three-qubit systems. For a three-qubit pure state $|\psi\rangle_{ABC}$, the following inequality holds

\begin{eqnarray}
{\tau}(|\psi\rangle_{A|BC})\leq {\tau}_a(\rho_{A|B})+{\tau}_a(\rho_{A|C})
\label{Eq13}
\end{eqnarray}
where ${\tau}(|\psi\rangle_{A|BC})$ is the tangle of the pure state $|\psi\rangle_{A|BC}$ between A and BC, and ${\tau}_a(\rho_{AB})= max\sum_{i}p_i{\tau}(|\psi\rangle_{AB})$ is the tangle of assistance of $\rho_{AB}=Tr_C|\psi\rangle_{ABC}\langle\psi| $ with the maximum taken over all possible pure-state decomposition $\rho_{AB}=\sum_{i}p_i|\psi_i\rangle_{AB}\langle\psi_i|$. This inequality was generalized into multi-qubit system
${\tau}_a(\rho_{A_1|A_2\ldots A_n})\leq \sum_{i=2}^n {{\tau}_a(\rho_{A_1|A_i})}$ for an arbitrary multi-qubit mixed state $\rho_{A_1A_2\ldots A_n}$ and its reduced density matrices $\rho_{A_1A_i}$ with $i=2,\ldots,n$. In Ref.\cite{Kimb:2012,Kim:2019,Lai:2021,Song:2019,Kim:2016,Song:2017,Kim:2012}, polygamy inequalities were also established for other entanglement measures.

For polygamy inequality beyond qubits, it was shown that von Neumann entropy can be used to establish a polygamy inequality of three-party quantum system\cite{Buscemi:2009}. We have $E(|\psi\rangle_{A|BC})\leq E_a(\rho_{A|B})+E_a(\rho_{A|C})$ for any three-party pure state $|\psi\rangle_{A|BC}$, where $E(|\psi\rangle_{A|BC})=S(\rho_A)=-Tr\rho_A ln\rho_A$ is
the von Neumann entropy of entanglement between A and BC, and $E_a(\rho_{AB})$ is the entanglement of assistance of $\rho_{AB}$ defined by $E_a(\rho_{AB})= max\sum_{i}p_i E(|{\psi}_i\rangle_{AB})$, where the maximization is taken over all possible pure state decompositions of $\rho_{AB}$. In Ref.\cite{Kimb:2012}, a general polygamy inequality of multipartite quantum entanglement was established for arbitrary-dimensional quantum states $\rho_{A_1A_2\ldots A_n}$. Recently, Kim\cite{Kim:2018} further proposed a class of weighted polygamy inequalities of multipartite entanglement in arbitrary-dimensional quantum systems.

\section{New definitions of MOE}
In this section we present some alternative methods to define MOE. The main problem with CKW inequalities is that their validity is not universal since several important measures of entanglement do not satisfy Eq.(\ref{Eq1}). In Ref.\cite{Lancien:2016}, Lancien \emph{et al} raise the following question: Should any entanglement measure be monogamous in a CKW-type sense? In fact, the summation in the right-hand side of Eq.(\ref{Eq1}) is only a convenient choice but not a necessity. For example, it has been shown that if $E$ does not satisfy the CKW monogamy inequalities, it is still possible to find a positive $\mu$ such that $E^{\mu}$ satisfies the Eq.(\ref{Eq1}). Inspired by this idea, one attempt is to replace Eq.(\ref{Eq1}) with the following generalized monogamy relation:
\begin{eqnarray}
E(\rho_{A|BC})\geq f(E(\rho_{A|B}),E(\rho_{A|C}))
\label{Eq14}
\end{eqnarray}
where $f:\mathbb{R}_+\times \mathbb{R}_+\rightarrow\mathbb{R}_+$ is a function independent on the dimension of the underlying Hilbert space, and it is continuous, and satisfies the condition $f(x,y)\geq max(x,y)$. This requirement comes from the fact that $E$ is an entanglement monotone which is nonincreasing under partial traces. The CKW-type monogamy inequality can be recovered for the particular choice $f(x,y)= x+y$. It has been proved that the entanglement of formation $E_F$ and the relative entropy of entanglement $E_R$, as well as their regularizations, cannot satisfy the new definition in Eq.(\ref{Eq14}). In addition, any additive entanglement measure which is geometrically faithful in the sense of being-lower bounded by a quantity with a sub-polynomial dimensional dependence on the antisymmetric state, cannot be monogamous. Nevertheless, we can recover the monogamy relation (\ref{Eq14}) if we allow the function to be dimension-dependent. For example, it has been shown that the non-trivial dimension-dependent monogamy relations can be established for $E_F$ and $E_R^{\infty}$ in any finite dimension.

Another approach to define MOE is given in terms of an equality, as opposed to the traditional monogamy inequality.  According to the  definition in Ref.\cite{Gour:2018}, a measure of entanglement $E$ is monogamous if for any $\rho_{ABC}\in S_{ABC}$ that satisfies
\begin{eqnarray}
E(\rho_{A|BC})=E(\rho_{AB})
\label{Eq15}
\end{eqnarray}
we have that $E(\rho_{AC})=0$. With respect to this definition, if the entanglement between system $A$ and the composite system $BC$ is as much as the entanglement that system $A$ shares with subsystem $B$, then it is left with no entanglement to share with $C$. If E satisfies Eq.(\ref{Eq1}), then any state $\rho_{A|BC}$ that satisfies the definition (\ref{Eq15}) must $E(\rho_{AC})=0$. Therefore, the condition in Eq.(\ref{Eq1}) is stronger than the definition in Eq.(\ref{Eq15}). This new definition is consistent with Eq.(\ref{Eq1}) and it has been shown that they are equivalent if and only if there exists $0\leq\mu\leq\infty$ such that

\begin{eqnarray}
E^{\mu}(\rho_{A|BC})\geq E^{\mu}(\rho_{AB})+E^{\mu}(\rho_{AC})
\label{Eq16}
\end{eqnarray}
for all $\rho_{ABC}\in S_{ABC}$ with fixed $dim\mathcal{H}_{ABC}=d<\infty$. It is to be noted that Eq.(\ref{Eq16}) is not a special case of Eq.(\ref{Eq14}) because the exponent factor $\mu$ depends on the dimension $d$, whereas the function $f$ defined in Eq.(\ref{Eq14}) is universal and does not depend on the dimension. By adopting the new definition of monogamy without inequalities, Guo and Gour \cite{Guo:2019} further proved the monogamy of EOF on mixed tripartite states.
\section{Concluding remarks and outlook}

The subject of MOE has attracted extensive research interest in the past two decades. In this review, we present the theoretical developments in the field of MOE, as well as some new definitions of MOE. Despite the rapid progress in recent years, there are still many challenging problems to be solved and we briefly list them as follows.

First, most previous studies of MOE are focussed on the multi-qubit systems. But our knowledge of MOE in the high-dimensional case is still very limited and there are few results on MOE for high-dimensional systems\cite{Choi:2015,2016:Tian,Kim:2021sr,Jin:2019qi,Jin:2019op}. In \cite{Kim:2008jpa} Kim \emph{et al.} proved that the $n$-qudit generalized W-class (GW) states satisfy the monogamy inequality in terms of the SC. Recently, Shi \emph{et al.} presented in \cite{Shi:2020} new monogamy and
polygamy relations for $n$-qudit generalized W-class
states and vacuum (GWV) states  in terms of the TqE. In \cite{Liang:2020,Lai:2021} the authors investigated the monogamy and polygamy relations for the GWV states in high-dimensional systems in terms of the R$\alpha$E. Except for squashed entanglement and one-way distillable entanglement, monogamy relations for various entanglement measures only hold for some special high-dimensional states. The difficulties are caused by the entanglement properties in higher-dimensional systems are hardly known so far and there is no analytical formula for calculating the high-dimensional entanglement measure. Thus, it is important to explore monogamy inequality for general high-dimensional states in terms of various entanglement measures.

Second, the validity of the traditional monogamy inequality is not universal and several important measures of entanglement do not satisfy Eq.(1). However, MOE has been mathematically proven to be a valid property of entanglement in the $n$-shareability sense\cite{Yang:2006}. In order to solve this problem, two new definitions of MOE have been proposed. One definition is to replace the right-hand side of Eq.(\ref{Eq1}) with a universal function $f$ independent of dimension $d$\cite{Lancien:2016}. This definition\cite{Lancien:2016} is somewhat artificial and some important entanglement measure such as EOF and relative entropy of entanglement cannot satisfy the new monogamy inequality. Another approach is to define MOE with an equality rather than inequality. It was shown that this definition is consistent with the traditional notion of MOE if the measure $E$ is replaced by $E^{\alpha}$ for some exponent $\alpha >0$. According to this new definition, EOF are monogamous on mixed tripartite systems. It supports that monogamy is a property of entanglement and not of some particular functions quantifying entanglement. Although the second definition of MOE seems more natural in physical, there is no mathematical proof of which definition is better, and we do not know whether there are entangled states that violate the second definition. Therefore, extensive efforts are still needed to investigate the relationship between these two definitions. Moreover, by adopting these new definitions, it is necessary to explore whether many important measures of entanglement are monogamous.

Third, different attempts have been made to construct a sharper version of monogamy inequality. In particular, Regula \emph{et al}\cite{Regula:2014} have proposed a set of SM inequalities in terms of concurrence. Although an extensive numerical evidence has been presented for four qubit systems, an analytical proof of SM conjecture is still desired. It would also be interesting to answer whether there are counterexamples that violate the SM inequality for more qubits. This conjecture can be further extended to negativity and SCREN for some classes of states. Future directions may include the study of SM inequalities for other entanglement monotones such as squashed entanglement.

In summary, we have reviewed the mathematical foundation of MOE but not include many problems concerning real physical phenomena, and monogamy is being considered in the study of these problems. For example, it was argued that the black hole evaporation is incompatible with our understanding of MOE\cite{Almheiri:2013}. Thus, it is desirable for us to have a sufficient understanding of monogamy further.

\acknowledgements
This work was supported by NSF-China under Grant Nos.11904071, the Anhui Provincial Natural Science Foundation under Grant Nos.1908085QA40.

\end{document}